# Exploring the high-pressure behavior of the three known polymorphs of $BiPO_4$: Discovery of a new polymorph

D. Errandonea[1,*], O. Gomis[2], D. Santamaría-Perez[1,3], B. García-Domene[1], A. Muñoz[4], P. Rodríguez-Hernández[4], S.N. Achary[5], A.K. Tyagi[5], and C. Popescu[6]

[1]*Departamento de Física Aplicada-ICMUV, MALTA Consolider Team, Universidad de Valencia, Edificio de Investigación, C/Dr. Moliner 50, Burjassot, 46100 Valencia, Spain*

[2]*Centro de Tecnologías Físicas, MALTA Consolider Team, Universitat Politècnica de Valencia, 46022 Valencia, Spain*

[3]*Earth Sciences Department, Univ. College London, UK*

[4]*Departamento de Física, Instituto de Materiales y Nanotecnología, MALTA Consolider Team, Universidad de La Laguna, La Laguna 38205, Tenerife, Spain*

[5]*Chemistry Division, Bhabha Atomic Research Centre, Trombay, Mumbai 400085, India*

[6]*CELLS-ALBA Synchrotron Light Facility, Cerdanyola, 08290 Barcelona, Spain*

## Abstract

We have studied the structural behavior of bismuth phosphate under compression. We performed x-ray powder diffraction measurements up to 31.5 GPa and *ab initio* calculations. Experiments were carried out on different polymorphs: trigonal (phase I) and monoclinic (phases II and III). Phases I and III, at low pressure (P < 0.2 – 0.8 GPa), transform into phase II, which has a monazite-type structure. At room temperature, this polymorph is stable up to 31.5 GPa. Calculations support these findings and predict the occurrence of an additional transition from the monoclinic monazite-type to a tetragonal scheelite-type structure (phase IV). This transition was experimentally found after the simultaneous application of pressure (28 GPa) and temperature (1500 K), suggesting that at room temperature the transition might by hindered by kinetic barriers. Calculations also predict an additional phase transition at 52 GPa, which exceeds the maximum pressure achieved in the experiments. This transition is from phase IV to an orthorhombic barite-type structure (phase V). We also studied the axial and bulk compressibility of $BiPO_4$. Room-temperature pressure-volume equations of state are reported. $BiPO_4$ was found to be more compressible than isomorphic rare-earth phosphates. The discovered phase IV was determined to be the less compressible polymorph of $BiPO_4$. On the other hand, the theoretically predicted phase V has a bulk modulus comparable with that of monazite-type $BiPO_4$. Finally, the isothermal compressibility tensor for the monazite-type structure is reported at 2.4 GPa showing that the direction of maximum compressibility is in the (0 1 0) plane at approximately 15º (21º) to the *a* axis for the case of our experimental (theoretical) study.

* Corresponding author, email: daniel.errandonea@uv.es

I.  **Introduction**

Bismuth phosphate ($BiPO_4$) is a multifunctional material with diverse applications. It is used as catalyst and photocatalyst, ion and humidity sensor, microwave dielectric, host for luminescent ions, and in the separation and immobilization of radioactive elements [1 – 9]. In contrast with related phosphates, $BiPO_4$ exhibits a rich structural polymorphism depending on preparation method [10]. In particular, three different crystal phases are known for $BiPO_4$, a trigonal structure (phase I) and two monoclinic structures which are obtained at low- (phase II) and high-temperature (phase III). The three structures are shown in Fig. 1. Phase II is stable under ambient conditions [10]. Its crystal structure belongs to space group (SG) *$P2_1/n$*, has four formula units per unit cell (Z = 4), and is isomorphic to the monazite structure [11]. Phase III is synthesized at high temperature [10] but can be recovered as a metastable phase at ambient conditions. The crystal structure of phase III is isomorphic to that of $SbPO_4$ and belongs to SG *$P2_1/m$* (Z = 2). Phase I is prepared by precipitation from an aqueous solution and the $H_2O$ molecules play a crucial role in the retention of this phase at ambient conditions [10]. The crystal structure of phase I belongs to SG *$P3_121$* (Z = 3). The structural relation among the three polymorphs has been discussed previously [10]. In particular, phases I and III consist of tetrahedral $PO_4$ groups and highly distorted eight-coordinated $BiO_8$ polyhedral units. In the case of phase II (monazite), the $BiO_8$ unit can be considered to be transformed into a $BiO_9$ polyhedron with eight Bi-O bonds within 2.36 to 2.70 Å and an additional long Bi-O bond at 3.02 Å.

Monazite-type oxides exist in nature. They are important accessory minerals in granitoids and rhyolites and are present in plutonic and metamorphic rocks. Therefore, the knowledge of the high-pressure (HP) structural behavior of monazite-type and related oxides is very relevant not only for technological applications, but also for mineral physics and chemistry as well as for petrology [12]. In this regard, monazite-type chromates [13, 14], vanadates [15, 16], and phosphates [17, 18] have been studied under compression. However, to the best of our knowledge, $SbPO_4$-type oxides have not been yet studied at HP. The same can be stated for phase I of $BiPO_4$. Under HP, monazite-type chromates have been found to undergo phase transitions at 3 GPa [13, 14] while in isomorphic vanadates the transitions are detected near 10 GPa [15]. In contrast, monazite-type phosphates are much more stable under compression. In particular, no phase transition is detected in $GdPO_4$, $EuPO_4$, and $NdPO_4$ up to 30 GPa and in $LaPO_4$ the onset of a structural transformation from the monazite-type to a barite-type structure occurs at 26 GPa [17]. On top of that, monazite $CePO_4$ does not undergo structural transitions up to 20 GPa, but its unit-cell parameters show an anomalous pressure behavior beyond 12 GPa [18]. Therefore, it is clear that further efforts are needed to elucidate the behavior under pressure of the different polymorphs of $BiPO_4$ and related phosphates.

Here, in order to improve the understanding of the structural properties of the three known polymorphs of $BiPO_4$ and to explore the possible occurrence of pressure-driven phase transitions, we have studied the HP behavior of phases I, II, and III of $BiPO_4$ by x-ray diffraction (XRD) up to 31.5 GPa. *Ab initio* calculations were also carried out, being obtained an excellent agreement with experiments. We have found that phase I and phase III transform into the monazite-type structure (phase II) at low pressure, 0.2 GPa and 0.8 GPa; respectively. Regarding phase II, calculations predict that a phase transition to a more dense scheelite-type structure (phase IV) should occur at 15 GPa. In contrast with this result, our room-temperature (RT) experiments found that phase II remains stable up to 31.5 GPa. However, upon the simultaneous application of pressure and temperature the new phase IV is obtained, suggesting that kinetic barriers could hinder the II→IV phase transition. The crystal structural details of the new phase have been determined. Moreover, calculations predict a phase transition from phase IV to an orthorhombic phase V occurring at 52 GPa, which is a pressure 20 GPa higher than the maximum experimental pressure. From our studies we obtained the axial and bulk compressibility for phases II and III as well as the isothermal RT P-V equation of state (EOS) for phases I to V. The reported results will be discussed in comparison with related phosphates.

**II. Experimental details**

Single-phase high-purity powders of $BiPO_4$ in phases I, II, and III were prepared by precipitation from an aqueous solution and subsequent treatments at different temperatures. Details on preparation method and sample characterization can be found elsewhere [10]. With the prepared samples we carried out four RT high-pressure experiments. A sample from phase III was compressed up to 28 GPa (run 1). A sample from phase I was pressurized up to 21 GPa (run 2). Two samples from phase II were studied up to 15.7 and 31.5 GPa (runs 3 and 4). Angle-dispersive XRD experiments were carried out using a diamond-anvil cell with diamond culets of 350 μm. The pressure chamber was a 100 μm hole drilled on rhenium gaskets pre-indented to 50 μm thickness. The studied samples were loaded in the pressure chamber together with a few W grains. The EOS of W [19] and the ruby fluorescence method [20] were used to determine pressure with an accuracy of 0.1 GPa. The presence in XRD patterns of Bragg peaks of W does not preclude the identification of the different crystal structure of $BiPO_4$. A 16:3:1 methanol-ethanol-$H_2O$ mixture was used a pressure transmitting medium [21]. Special care was taken to occupy only a small fraction on the pressure chamber with the loaded samples to reduce the possibility of sample bridging between the diamond anvils [22, 23]. In situ HP XRD measurements were carried out at MSPD beamline of ALBA synchrotron [24] with the exception of run 3 which was performed using an Xcalibur diffractometer [25]. At ALBA the incident monochromatic beam of wavelength 0.4246 Å was focused down to a 10 μm × 15 μm spot using Kirkpatrick-Baez mirrors and a Rayonix CCD detector was used to collect XRD patterns. In run 3, XRD patterns were obtained on a 135-mm Atlas CCD detector using $K_{\alpha 1}$:$K_{\alpha 2}$ Mo radiation being the x-ray beam collimated to a diameter of 300 μm. The two dimensional diffraction images collected in runs 1, 2, and 4 at ALBA were integrated with the FIT2D software [26] whereas the two dimensional diffraction images collected in run 3 were integrated with the CrysAllis software [27]. Structural analysis was performed with PowderCell [28] and GSAS [29]. In run 1, after finalizing the compression cycle at 28 GPa, a thermal annealing was carried out searching for the theoretically predicted new polymorph of $BiPO_4$. For the thermal treatment we used a laser-heating set-up equipped with a 100 W fiber laser (λ = 1064 nm) [30]. The sample was heated to 1500 K for 2 minutes and then quenched. Temperature was calculated by fitting a Planck function to the measured thermal emission spectrum of the sample [31].

**IIII. Theoretical methods**

*Ab initio* total energy simulations have been performed within the density functional theory (DFT) framework as implemented in the Vienna *ab initio* simulation package (VASP) [32]. VASP performs structural calculations with the plane wave pseudo-potential method. In our study, the set of plane waves used was extended up to a kinetic energy cutoff of 520 eV to achieve highly converged results within the projector-augmented-wave scheme. In addition, the exchange-correlation energy was taken in the generalized gradient approximation (GGA) with the revised Perdew–Burke–Ernzerhof (PBESOL) [33] prescription which works better for $BiPO_4$ than the local density approximation (LDA) [10]. Moreover, we used dense special point grids appropriate to each structure to sample the Brillouin zone, ensuring a high convergence (1–2 meV) per formula unit in the total energy of each structure as well as a precise determination of the forces on the atoms. At each selected volume, the structures were fully relaxed to their equilibrium configurations through the calculation of the forces on the atoms and the stress tensor. In the relaxed equilibrium configuration, the forces were < 0.006 eV Å$^{-1}$, and the deviation of the stress tensor from a diagonal hydrostatic form was < 0.1 GPa. Consequently, our calculations provide a set of accurate energy, volume, and pressure ($E$, $V$, $P$) values that can be fitted using an EOS in order to obtain the equilibrium volume ($V_0$), bulk modulus ($B_0$), and its pressure derivatives ($B_0$' and $B_0$''). From the calculated data it is also possible to determine the thermodynamically most stable structure at different pressures [34]. With this aim, in addition to the three known polymorphs of BiPO$_4$ [10], we have also included in the simulations the scheelite-type (SG $I4_1/a$ Z = 4) and the barite-type (SG $Pnma$, Z = 4) structures, which were previously observed as HP phases in related oxides [17, 35].

### IV. Results and discussion

#### A. High-pressure x-ray diffraction

Fig. 2 shows XRD patterns measured in run 1 starting from phase III up to 22 GPa. In this figure, a Bragg peak associated to W can be easily identified since it has a different pressure evolution than those of the sample. In this run, we found that the patterns obtained from ambient pressure up to 0.6 GPa can be unequivocally assigned to the SbPO$_4$-type structure. This is illustrated in Fig. 2 by the XRD patterns measured at 0.1 and 0.6 GPa. For the first one the residuals of the structural refinement are shown. The $R$-factors of the refinement are $R_p$ = 3.01% and $R_{wp}$ = 6.05%. The unit-cell parameters determined at 0.1 GPa are $a$ = 4.871(5) Å, $b$ = 7.081(7) Å, $c$ = 4.696(5) Å, and $\beta$ = 96.17(9)°. When pressure reaches 0.8 GPa new Bragg peaks emerge which are identified by asterisks in Fig. 2. Upon compression, these and other extra peaks gradually grow in intensity and simultaneously the peaks assigned to phase III gradually vanish. These changes can be ascribed to the onset of a phase transition, coexisting phase III and the HP phase from 0.8 to 3.0 GPa. The HP phase appears as a single phase at 4.2 GPa and can be assigned to the monazite-type structure (phase II). The structural assignment for the HP phase is supported by Rietveld refinements. The residuals of the refinement made for the data collected at 4.2 GPa are shown in Fig. 2. The *R*-factors of the refinement are $R_p$ = 3.85% and $R_{wp}$ = 6.83%. The unit-cell parameters of monazite-type $BiPO_4$ at 4.2 GPa are *a* = 6.646(7) Å, *b* = 6.876(7) Å, *c* = 6.407(5) Å, and $\beta$ = 103.3(1)º. Pressure release from phase II at 4.2 GPa shows that the observed transition is not reversible. Upon further compression up to 28 GPa no further phase transitions are found. This can be seen in the XRD patterns measured at 22 GPa (Fig. 2) and 28 GPa (Fig. 3). As we will show in the next section, this result seems to be in conflict with our calculations, which predict that phase II should undergo a phase transition beyond 15 GPa. One possible reason for this is the presence of large kinetic barriers which hinder the occurrence of the phase transition [23]. To check this hypothesis we carried out a laser-heating annealing of our sample at 1500 K. After this treatment we found that $BiPO_4$ was crystallized in a different structure which can be assigned to the HP phase predicted by the calculations (phase IV). This is shown in Fig. 3. Rietveld refinements of the XRD pattern collected from phase IV indicated that they are consistent with a scheelite-type structure (see phase IV in Fig. 1) with *a* = 4.66(1) Å and *c* = 11.07(2) Å at 28 GPa, being the *R*-factors of the refinement $R_p$ = 2.74% and $R_{wp}$ = 6.28%. This result agrees with the theoretically predicted structure. Upon pressure release a mixture of phase II and IV is recovered at 0.1 GPa.

We will summarize now the results obtained in the other experimental runs. XRD patterns are not shown to avoid redundancies. In run 2 we used a sample in which the trigonal phase I was observed before sample loading. Details on the characterization of this sample are given in Ref. 10. However, after loading trigonal $BiPO_4$ in a DAC, the monoclinic phase II was found after the first compression step at a pressure smaller than 0.2 GPa. This experiment was repeated in a second DAC loaded with the same sample (phase I) and the same result was obtained (phase II at 0.2 GPa). We believe that this fact could be due to desorption of $H_2O$ molecules by the effect of pressure. It is also consistent with the fact that, according with calculations, phase I is thermodynamically the less favored polymorph and phase II is the most stable one (see next section). Upon further compression in run 2, phase II is retained up to the highest pressure covered by the experiments in agreement with the results of run 1. Upon decompression phase II is recovered, so the I→II transition is not reversible. Finally, runs 3 and 4 were carried out directly on samples with the monazite-type structure. We found that at RT this phase II is stable from ambient up to the highest pressure reached in our study, 31.5 GPa.

**B. *Ab initio* calculations**

For our theoretical study of the structural stability of $BiPO_4$ at HP, we have taken into consideration previous results obtained in $ABO_4$ ternary oxides and the packing-efficiency criterion [36]. In addition to the three known polymorphs of $BiPO_4$, we have studied the relative stability of two HP candidate structures using the calculation method outlined above. These two structures are isomorphic with scheelite [37] and barite [38]. They are represented in Fig. 1 as phases IV and V, respectively. Fig. 4 shows the total energy versus volume and enthalpy difference versus pressure curves for the different structures that have been considered. The monazite-type structure (phase II) is the one with the lowest energy and enthalpy at ambient pressure.

Therefore, monazite is the stable structure of $BiPO_4$. The calculated structural parameters for $BiPO_4$ (phases I, II, and III) at ambient pressure have been reported elsewhere [10] agreeing very well with the experimental results. Upon compression calculations predict the occurrence of a monazite-to-scheelite phase transition at 15 GPa. This is a first-order transition that involves a large volume collapse ($\Delta V/V = -9$ %) and according with the literature there is a large kinetic barrier associated to it [39]. This could explain why in the experiments we did not found the transition at RT, but detected it only after heating at 1500 K. The calculated structural parameters of phase IV at 27.6 GPa are given in Table I. Additionally, in the simulations we have found that at 52 GPa the barite-type structure (phase V) becomes thermodynamically more stable than any other structure in $BiPO_4$. The scheelite-barite transition involves also a volume collapse ($\Delta V/V = -2$ %); i.e. it is a first-order transition. The potential appearance of a barite-type structure at HP in $BiPO_4$ is consistent with the results found in LaPO4 [17]. In our experiments, phase V has not been observed, but since the calculated transition pressure is 20 GPa higher than the maximum pressure achieved in the experiments, it would not be surprising that phase V could be found in future experiments beyond 50 GPa, in special if thermal annealing is used to overcome kinetic barriers. The calculated structural details of phase V at 61 GPa are given in Table I. An interesting fact to remark is that in the five polymorphs of $BiPO_4$ the P atoms are four coordinated to oxygen atoms. In contrast, Bi is 8 (or 9) coordinated in the four phases found in the experiments, but 12 coordinated in the barite-type structure. Thus the theoretically predicted IV-V transition involves more important atomic rearrangements than I-II, III-II, and II-IV transitions. In addition, the scheelite-type structure is the most symmetric among the five structures studied having perfect regular $PO_4$ tetrahedra and a $BiO_8$ dodecahedra with only two different Bi-O distances. This structure has been found before as a post-monazite structure in TbPO$_4$ and other phosphates that crystallize at ambient conditions in the zircon structure [39, 40]. This consistency between different studies supports the existence of the monazite-to-scheelite phase transition in BiPO$_4$.

### C. Room-temperature equations of state

From our experiments we extracted the pressure evolution of the unit-cell parameters for phases III and II of BiPO$_4$. The results are summarized in Figs. 5 and 6 and compared with calculations. The experimental and calculated parameters show a quite good agreement. In SbPO$_4$-type BiPO$_4$ (Fig. 5), the compressibility of the three axes is similar up to 3 GPa. The three axial compressibilities determined from the experiments are $\sim 4 \times 10^{-3}$ GPa$^{-1}$. In addition, the $\beta$ angle decreases upon compression. In monazite-type BiPO$_4$ (Fig. 6) the compression is not isotropic, being $a$ the most compressible axis. As a consequence of it, $a$ becomes very similar to $c$ at 31.5 GPa. On top of that, as in phase III, the $\beta$ angle of phase II also decreases under compression. The response to pressure of monazite BiPO$_4$ is similar to that found in other isomorphic phosphates [17]. As in most of them, no unusual changes on the pressure dependence of the unit-cell parameters have been detected in BiPO$_4$. Thus, the anomalies found in CePO$_4$ could be either an experimental artifact [22, 23] or be caused by an isomorphic second-order phase transition as the one occurring in monazite-type PbCrO$_4$ [13, 14]. Thus CePO$_4$ deserves to be systematically studied in future works.

A detailed discussion of the axial compressibilities of monazite BiPO$_4$ by means of the compressibility tensor will be made in the next section. Here we will concentrate on the RT EOS of the different phases obtained with the EosFit7c package [41]. For phase II, the evolution of the volume with pressure can be well described by a 3$^{rd}$ order Birch–Murnaghan (BM3) EOS [42] (see Fig. 6). The obtained EOS parameters are given in Table II. Again theory and experiments compare quite well with each other.

From the determined bulk modulus at zero pressure, $B_0$, it can be concluded that monazite-type $BiPO_4$ is more compressible than most rare-earth phosphates [17], but has a $B_0$ comparable with that of $CePO_4$ [18]. A visual indication of the quality of the EOS fit is provided in the inset of Fig. 6 where the normalized pressure ($F$) is plotted versus the Eulerian strain ($f_E$) [43]. There it can be seen that the $F$–$f_E$ relation lies on a straight line with a positive slope, indicating that the experimental data are adequately described by a BM3 EOS. From a linear least-squared fit to the $F$–$f_E$ data we obtained $B_0 = 101(1)$ GPa and $B_0' = 5.7(9)$ [43]. These values are consistent with those obtained from the EOS fit to the experimental results.

For phase III we have also made an EOS fit to the P-V data. In this case, since we have only five experimental data points we used a $2^{nd}$ order Birch-Murnaghan (BM2) EOS. The EOS parameters obtained are given in Table II. Again the agreement between experiment and theory is quite good. Since $B_0$ and $B_0'$ are correlated parameters, to compare phase II with phase III we have also fit the results obtained for phase II with a BM2 EOS. The obtained results are shown in Table II. There it can be seen that phase III is much more compressible than phase II. In particular, the bulk modulus at zero pressure, $B_0$, of $SbPO_4$-type $BiPO_4$ is comparable to that of $CrVO_4$-type orthophosphates [35].

In Table II, we also give the theoretical EOS of phases I, IV, and V. Phase I is the most compressible structure. A higher compressibility among orthophosphates is only found in quartz-like and berlinite-type phosphates [44, 45]. Regarding phase IV, we found that it is less compressible than the three ambient pressure polymorphs. The increase of the bulk modulus in phase IV is related with the increase of the packing efficiency in it. It is also important to note here that if a $4^{th}$ order EOS is used to fit the theoretical and experimental results, none of the four structures shows anomalous positive values for the second pressure derivative of the bulk modulus ($B_0''$). This fact implies that the rate at which all phases become stiffer decreases with increasing pressure. On the other hand, from our calculations we conclude that in the five structures (found or predicted) in $BiPO_4$ the Bi-O bonds are much more compressible than the P-O bonds. Consequently they account for most of the volume reduction.

Regarding the scheelite phase, we would like to note that in spite of being the less compressible phase of $BiPO_4$, it still shows a bulk modulus (151 GPa) considerably smaller than the HP scheelite phases found in other trivalent metal phosphates [17] like scheelite-type $ScPO_4$ in which $B0 = 376$ GPa. Since the P-O bonds are known to be uncompressible [17, 39, 40], a possible explanation for this fact might be that Bi-O bonds are more compressible than Sc-O bonds. This hypothesis is consistent with the fact that all the know polimorphs of $Bi_2O_3$ has a bulk modulus ($B_0 < 100$ GPa) [46] which is considerable smaller than the bulk modulus of $Sc_2O_3$ ($B_0 = 188$ GPa) [47].

Let us now comment here on the potential hardness of scheelite-type $BiPO_4$. Several authors have demonstrated an empirical correlation between hardness and bulk modulus of materials [48]. By applying this correlation to scheelite-type $BiPO_4$, the Vickers hardness is estimated to be approximately 8 GPa, a value similar to that experimentally determined for related scheelite-type phosphates [49]. This is an interesting property which can be useful for technological applications since scheelite-type $BiPO_4$ can be potentially recovered as a metastable phase at ambient conditions. It would be also interesting to explore in the future the use of innovative preparation techniques of $BiPO_4$ nanoparticles [50] to explore the potential preparation of scheelite-type $BiPO_4$ nanostructures. Their preparation can be useful to enhance the photocatalytic activity of $BiPO_4$ since, for a given compound, scheelite phases are known to have a smaller electronic band gap than monazite phases [51].

Another fact to highlight is related to the EOS of phase V. According with our calculations, the barite-type phase V has a bulk modulus smaller than scheelite-type phase IV. This fact is apparently in contradiction with the density increase associated to the scheelite-barite transition. However, similar phenomena, although not common, have been also reported in other pressure-driven phase transitions; e.g. the B1–B2 transformation in alkali halides [52]. This phenomenon was assigned to a bond strength decrease associated to the phase transition. This hypothesis is plausible in our case where the coordination increase of Bi associated to the scheelite-barite transition is accompanied by an increase of the average Bi-O interatomic distance.

**D. Isothermal compressibility tensor**

The isothermal compressibility tensor, $\beta_{ij}$, is a symmetric second rank tensor which relates the state of strain of a crystal to the change in pressure that induced it [53]. This tensor for a monoclinic crystal has as coefficients:

$$\beta_{ij} = \begin{pmatrix} \beta_{11} & 0 & \beta_{13} \\ 0 & \beta_{22} & 0 \\ \beta_{13} & 0 & \beta_{33} \end{pmatrix}$$

Using the IRE (Institute of Radio Engineers) convention for the orthonormal basis for the tensor: $e_3 \| c$, $e_2 \| b^*$, $e_1 \| e_2 \times e_3$, we have obtained the isothermal compressibility tensor coefficients for phase II of $BiPO_4$ at a pressure of 2.4 GPa. The tensor has been obtained using the linear Lagrangian approximation with the equations given in Ref. [54] and with the infinitesimal Lagrangian approximation as implemented in the Win-Strain package [55]. For the case of the linear Lagrangian approximation a linear fit of the unit-cell parameters was carried out in the pressure range 0-4.8 GPa where the unit-cell parameters behaviour was found to be linear. Table III reports the values of the lattice parameters at 1 atm and their pressure derivatives from the linear fits, which are used in the linear Lagrangian approximation, for the case of our

experimental and *ab initio* calculated data. Table III also includes the $\beta_{ij}$ coefficients of the isothermal compressibility tensor with the two approximations used. It can be observed that the agreement between the experimental and calculated data is quite good. On the other hand, the $\beta_{ij}$ coefficients obtained with the linear Lagrangian approximation agree, within the experimental uncertainties, with those obtained with the infinitesimal Lagrangian approximation as expected according to our small-strains assumption. The eigenvalues and eigenvectors for the isothermal compressibility tensor are reported in Table III. Taking into account the eigenvalues, it is found that for our experiments with the linear Lagrangian approximation, the maximum, intermediate and minimum compressibilities are $3.95(21) \times 10^{-3}$, $2.33(12) \times 10^{-3}$, and $1.84(21) \times 10^{-3}$ GPa$^{-1}$, respectively; whereas for the case of our calculations the obtained values for the compressibilities are $3.91 \times 10^{-3}$, $2.56 \times 10^{-3}$ and $1.74 \times 10^{-3}$ GPa$^{-1}$. These results indicate that around 50% of the total compression over the pressure range 0 - 4.8 GPa, is being accommodated along the direction of maximum compressibility. Taking into account the eigenvector $ev_1$, the major compression direction occurs in the (0 1 0) plane at the given angle $\Psi$ (see Table III) to the *c* axis (from *c* to *a*). Note that the direction of maximum compressibility, taking into account the value of $\beta_0$, is at approximately 15º (21º) to the *a* axis for the case of our experimental (theoretical) data. The direction of intermediate compressibility (see eigenvector $ev_2$) is along the *b* axis, and the direction of minimum compressibility (see eigenvector $ev_3$) is in the (0 1 0) plane perpendicular to the direction of maximum compressibility. To conclude, we note that the isothermal compressibility tensor has not been obtained for phase III of BiPO$_4$ because of the small stability range of this monoclinic phase in which we have only structural information for the pure phase from our experiments at two different pressures, 0.1 and 0.6 GPa.

**IV. Concluding remarks**

In this work we reported an experimental and theoretical study of the structural stability of the different polymorphs of $BiPO_4$ under compression. XRD experiments together with calculations have allowed us to determine that phases I and III of $BiPO_4$ transform into the monazite-type polymorph (phase II) at very low pressure. Both phase transitions are irreversible. In addition, in our RT experiments it is found that phase II remains stable up to 31.5 GPa. In contrast, calculations predict a monazite → scheelite (phase IV) → barite (phase V) structural sequence under pressure. The monazite-scheelite transition was found experimentally upon the combined application of pressure and temperature, indicating that kinetic barriers may have hindered its finding at RT. The phase IV found experimentally has the scheelite crystal structure predicted by calculations. The barite phase V is theoretically predicted to occur at a pressure 20 GPa larger than the maximum pressure covered by our experiments. The RT equations of state of the different phases are also reported and discussed. Furthermore, the isothermal compressibility tensor is given for phase II at 2.4 GPa and its eigenvalues and eigenvectors are obtained providing information about the directions of maximum, intermediate and minimum compressibilities. Finally, the reported results are discussed in comparison with the HP structural behavior of related phosphates. We hope the results here reported will stimulate additional HP studies in $BiPO_4$ and related oxides.

**Acknowledgements**

Research supported by the Spanish government MINECO under Grant No: MAT2013-46649-C4-1/2/3-P and by Generalitat Valenciana under Grants Nos: GVA-ACOMP-2013-1012 and GVA-ACOMP/2014/243. B.G.-D. thanks the financial support from MEC through FPI program. Experiments were performed at MSPD beamline at ALBA Synchrotron Light Facility with the collaboration of ALBA staff.

**Table I:** Calculated atomic positions in scheelite-type $BiPO_4$ (top) at 27.6 GPa ($a$ = 4.715 Å and $c$ = 11.064 Å) and barite-type (bottom) $BiPO_4$ at 61.0 GPa ($a$ = 7.503 Å, $b$ = 4.715 Å, and $c$ = 6.106 Å). Wyckoff positions are indicated.

| Atom | x | y | z |
| --- | --- | --- | --- |
| Bi (4b) | 0 | 0.25 | 0.625 |
| P (4a) | 0 | 0.25 | 0.125 |
| O (16f) | 0.24030 | 0.12805 | 0.04842 |
| Bi (4c) | 0.17931 | 0.25 | 0.17797 |
| P (4c) | 0.06796 | 0.25 | 0.68685 |
| $O_1$ (4c) | 0.39975 | 0.25 | 0.94266 |
| $O_2$ (4c) | 0.23803 | 0.25 | 0.55725 |
| $O_3$ (8d) | 0.07841 | 0.50231 | 0.83470 |

**Table II:** EOS parameters for the different phases of BiPO$_4$. The last two columns indicate the EOS type used and if results come from experiment (E) or theory (T).

| Phase | $V_0$ [Å$^3$] | $B_0$ [GPa] | $B_0'$ | EOS | |
|---|---|---|---|---|---|
| I | 273.66 | 64.4 | 5.53 | BM3 | T |
| II | 295.68 | 112.14 | 4.44 | BM3 | T |
| II | 294.46 | 119.80 | 4 | BM2 | T |
| II | 295.4(3) | 99(2) | 5.8(3) | BM3 | E |
| II | 294.2(2) | 117(1) | 4 | BM2 | E |
| III | 160.42 | 88.18 | 2.92 | BM3 | T |
| III | 160.01 | 84.8 | 4 | BM2 | T |
| III | 160.0(2) | 78(4) | 4 | BM2 | E |
| IV | 281.13 | 151.26 | 4.85 | BM3 | T |
| V | 285.62 | 120.48 | 4.27 | BM3 | T |

**Table III:** Lattice parameters at 1 atm, their pressure derivatives, the isothermal compressibility tensor coefficients, $\beta_{ij}$, and their eigenvalues, $\lambda_i$, and eigenvectors, $ev_i$, for monazite-BiPO$_4$ (phase II) at 2.4 GPa. The results are given using the linear Lagrangian and the infinitesimal Lagrangian methods with data from our experiments and our theoretical calculations.

| Method | Linear Lagrangian | | Infinitesimal Lagrangian | |
|---|---|---|---|---|
| | Experiment | Theory | Experiment | Theory |
| $a_0$ (Å), d$a$/d$P$ (Å·GPa$^{-1}$) | 6.756(1), -0.0256(7) | 6.7554, -0.0246(5) | | |
| $b_0$ (Å), d$b$/d$P$ (Å·GPa$^{-1}$) | 6.940(1), -0.0162(8) | 6.9553, -0.0178(4) | | |
| $c_0$ (Å), d$c$/d$P$ (Å·GPa$^{-1}$) | 6.472(1), -0.0151(9) | 6.47054, -0.0158(2) | | |
| $\beta_0$ (°), d$\beta$/d$P$ (°·GPa$^{-1}$) | 103.65(5), -0.086(4) | 103.954, -0.1033(8) | | |
| $\beta_{11}$ (10$^{-3}$ GPa$^{-1}$) | 3.45(18) | 3.21 | 3.50(13) | 3.20 |
| $\beta_{22}$ (10$^{-3}$ GPa$^{-1}$) | 2.33(12) | 2.56 | 2.28(8) | 2.57 |
| $\beta_{33}$ (10$^{-3}$ GPa$^{-1}$) | 2.34(14) | 2.44 | 2.16(8) | 2.44 |
| $\beta_{13}$ (10$^{-3}$ GPa$^{-1}$) | -0.90(19) | -1.02 | -0.96(13) | -1.03 |
| $\lambda_1$ (10$^{-3}$ GPa$^{-1}$) | 3.95(21) | 3.91 | 4.01(14) | 3.92 |
| $ev_1$ ($\lambda_1$) | (0.873, 0, -0.487) | (0.822, 0, -0.570) | (0.886, 0, -0.463) | (0.821, 0, -0.571) |
| $\lambda_2$ (10$^{-3}$ GPa$^{-1}$) | 2.33(12) | 2.56 | 2.28(8) | 2.57 |
| $ev_2$ ($\lambda_2$) | (0, 1, 0) | (0, 1, 0) | (0, 1, 0) | (0, 1, 0) |
| $\lambda_3$ (10$^{-3}$ GPa$^{-1}$) | 1.84(21) | 1.74 | 1.66(14) | 1.72 |
| $ev_3$ ($\lambda_3$) | (0.487, 0, 0.873) | (0.570, 0, 0.822) | (0.463, 0, 0.886) | (0.571, 0, 0.821) |
| $\Psi$ (°)[a] | 119(4) | 124.7 | 117.6(2.5) | 124.8 |

[a] The major compression direction occurs in the (0 1 0) plane at the given angle $\Psi$ to the $c$ axis (from $c$ to $a$).

**Figure captions**

**Figure 1: (color online)** Schematic view of the crystal structure of different polymorphs of BiPO$_4$. Bi atoms: purple, O atoms: red, P atoms: gray. The coordination polyhedra are shown.

**Figure 2:** XRD patterns collected in run 1 up to 22 GPa. At selected pressures the measured pattern (dots) is shown together with the calculated profile and residuals (lines). The ticks indicate the position of Bragg reflections.

**Figure 3: (color online)** XRD patterns measured at 28 GPa before and after laser-heating annealing and at 0.1 GPa after decompression. The measured patterns (dots) are shown together with the calculated profiles and residuals (lines). The ticks indicate the position of Bragg reflections.

**Figure 4: (color online)** (a) Energy vs volume and (b) enthalpy difference vs pressure plots for the different structures of BiPO$_4$. To facilitate comparison, volumes have been normalized assuming 4 formula units for all the structures. The structures have been named as I, II, III, IV, and V following the main text.

**Figure 5: (color online)** Pressure dependence of unit-cell parameters and volume for phase III. Symbols: experiments. Solid lines: fitted BM2 EOS for volume and guides to the eye for *a*, *b*, *c*, and *β*. Dashed lines: calculations.

**Figure 6: (color online)** Pressure dependence of unit-cell parameters and volume for phase II. Symbols: experiments, ● run 1, ▲ run 2, ■ run 3, and ▼ run 4. Solid lines: fitted BM3 EOS for volume and guides to the eye for *a*, *b*, *c*, and β. Dashed lines: calculations. The inset shows the normalized pressure vs Eulerian strain ($F$–$f_E$) plot.

**Figure 1**

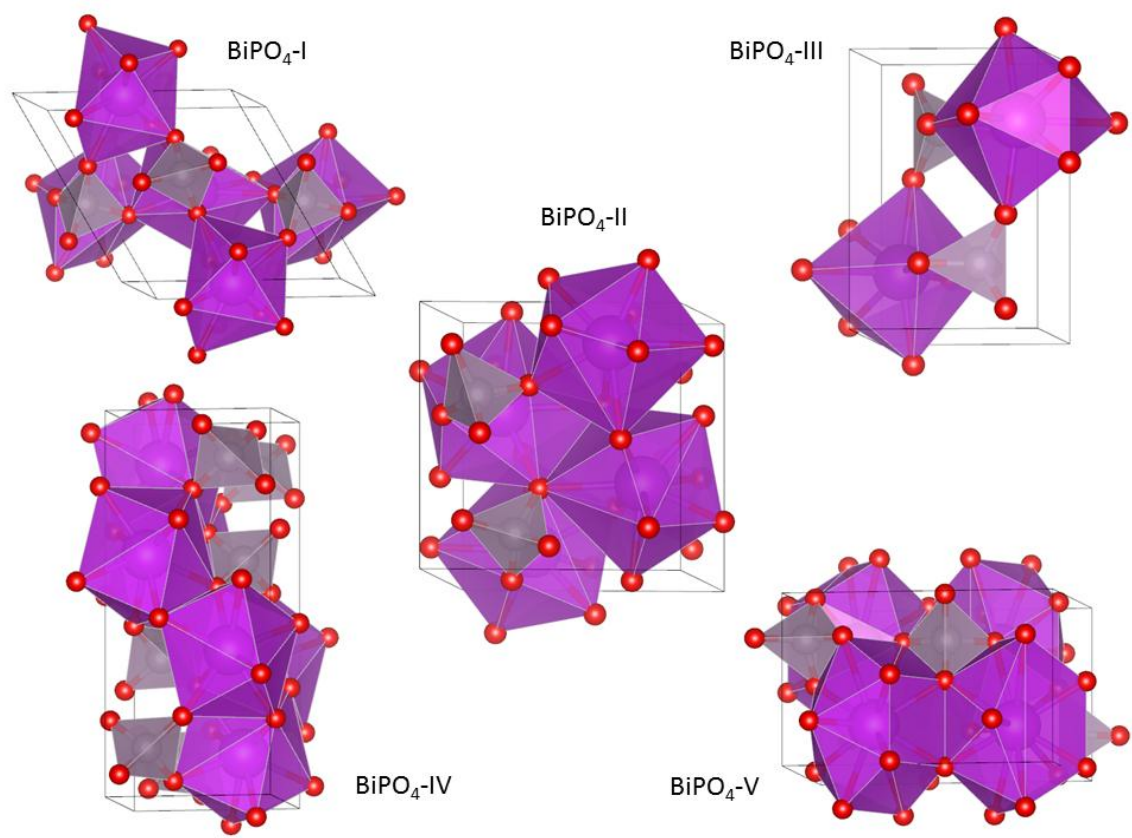

**Figure 2**

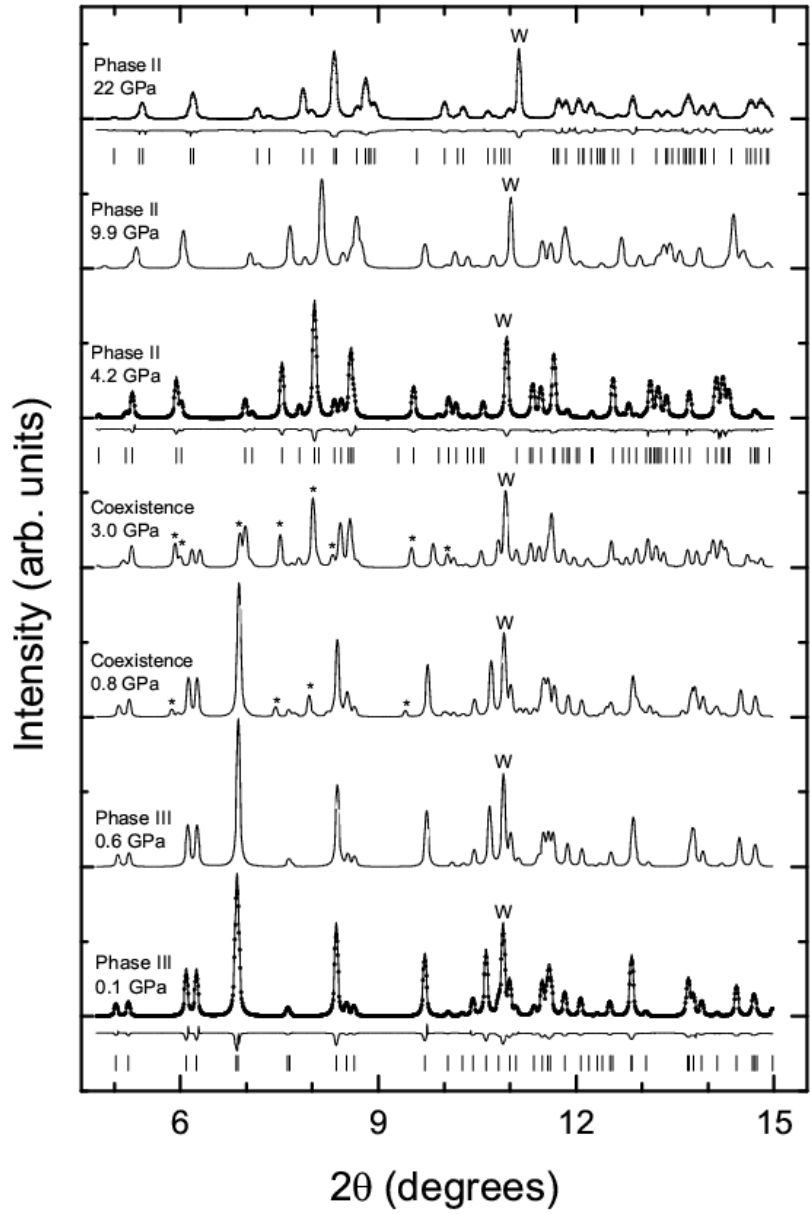

**Figure 3**

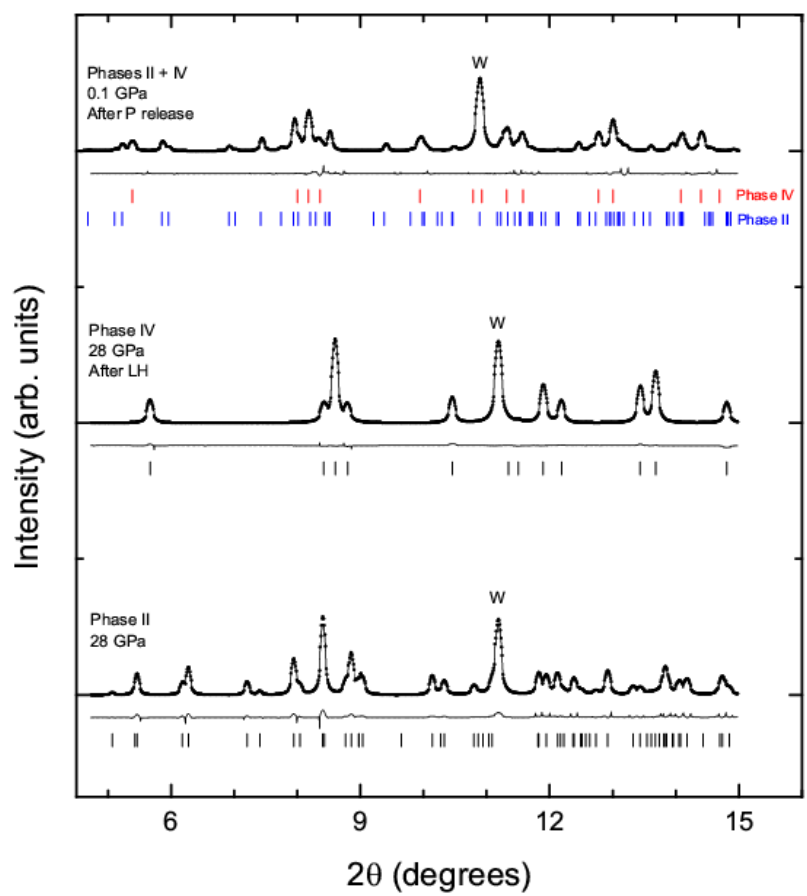

**Figure 4**

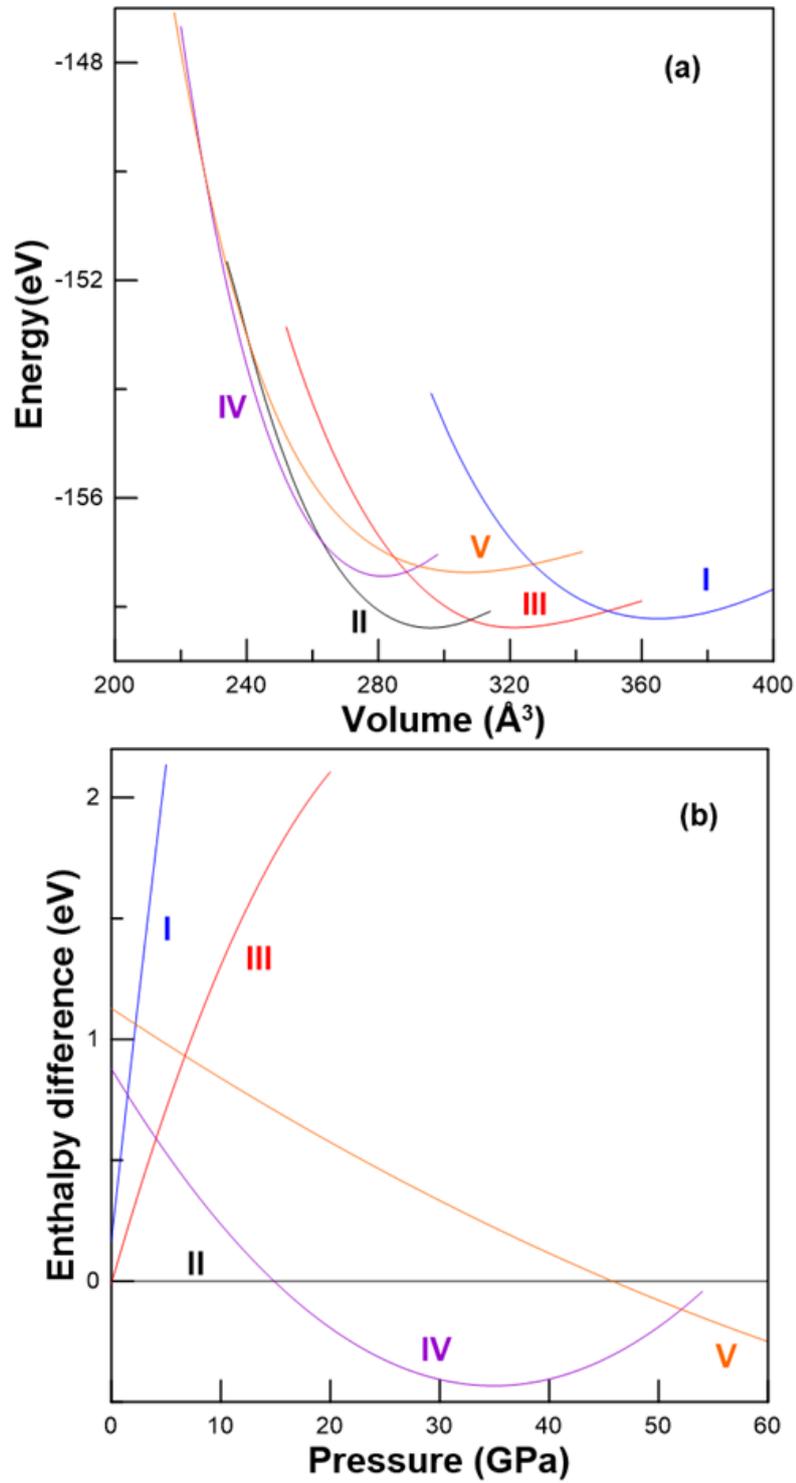

**Figure 5**

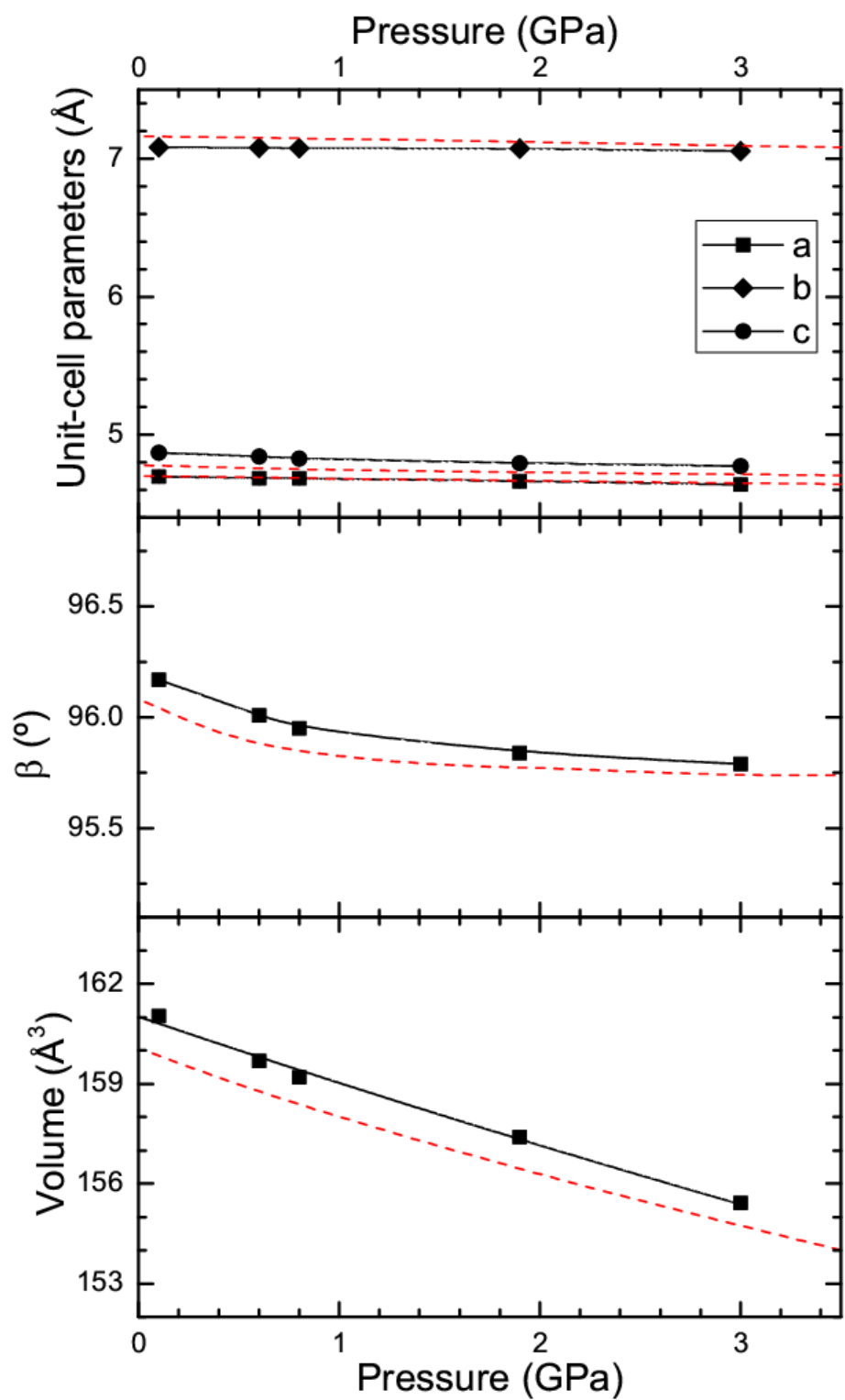

**Figure 6**

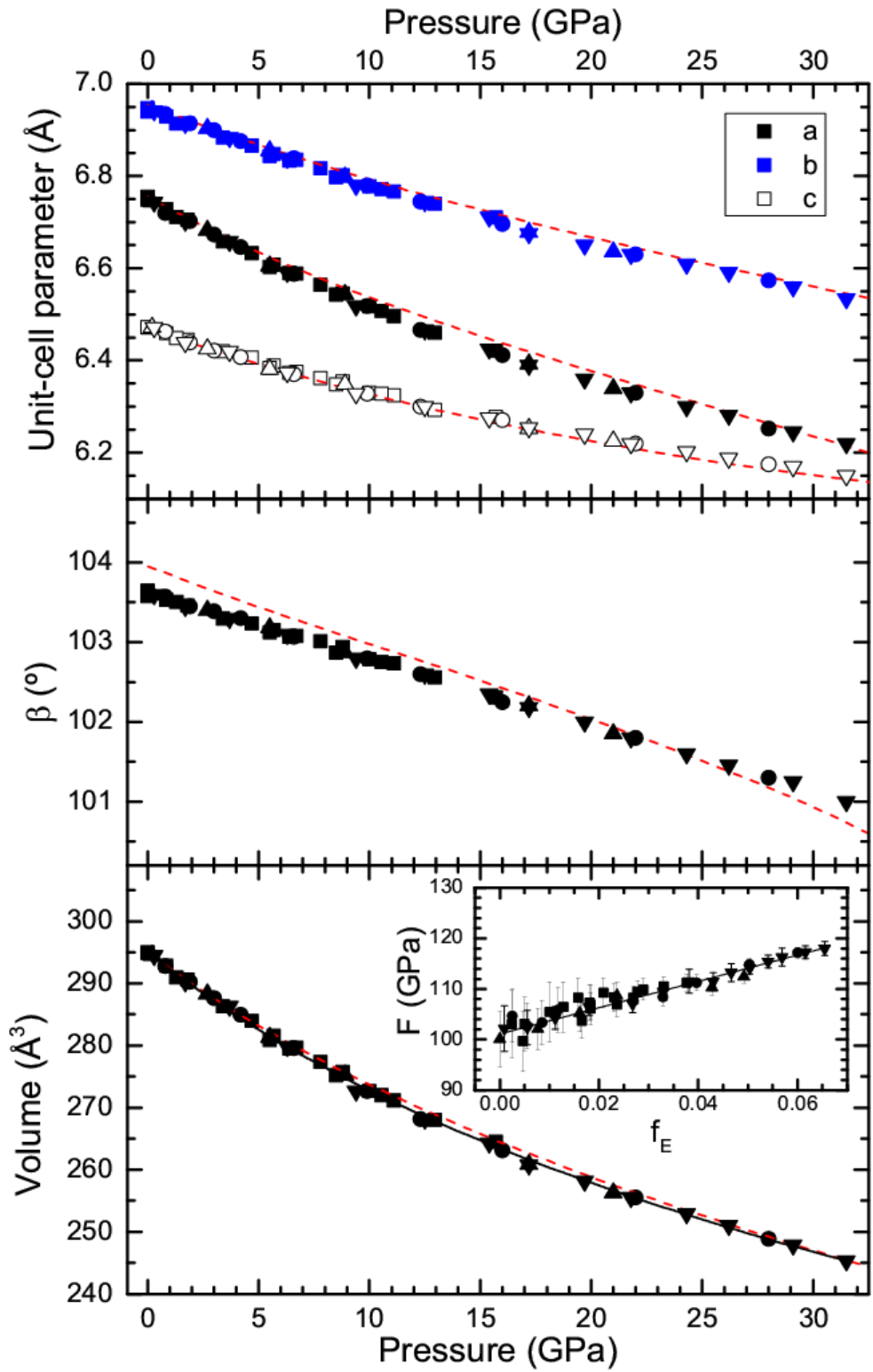